\documentclass{aa}

\usepackage{epsfig}			

\usepackage{txfonts}             

\begin{document}

   \title{An optical absorption study of the Helix  Planetary 
   Nebula}
  \subtitle{Na\,{\sc i} and Ca\,{\sc ii} lines, and a search for diffuse bands}
   \author{N. Mauron\inst{1} \and T.R. Kendall\inst{2} }
   \offprints{N.Mauron}

   \institute{ Groupe d'Astrophysique, UMR 5024 CNRS, Case CC72, Place Bataillon, 
	     F-34095 Montpellier Cedex 05, France\\
             \email{mauron@graal.univ-montp2.fr}
      \and 
     Laboratoire d'Astrophysique, Observatoire de Grenoble, 
   Universit\'{e} Joseph Fourier, BP 53, F-38041 Grenoble Cedex 09, France
   \email{tkendall@obs.ujf-grenoble.fr}
\thanks{Based on observations carried out with the European 
	     Southern Observatory VLT/UVES, program 70.C-0100}}

   \date{Received xxx; accepted xxx}

   \abstract {We present the first results of an optical absorption study 
 of NGC\,7293, the Helix planetary nebula (PN), designed to search definitively  
 for diffuse band (DIB) absorptions perhaps arising in the carbon-rich 
 circumstellar matter of the PN. We used the VLT-UVES spectrograph to achieve 
 high resolution (R=50000) spectra of 8 stars located angularly close to and 
 behind the Helix. These targets were selected through their photometric UBV 
 or 2MASS properties, permitting derivation of spectral types (mainly F-G dwarfs) 
 and distance that place several of them far (700--1500\,pc) beyond the Helix 
 (210\,pc). Through a detailed analysis of the Na\,{\sc i} and Ca\,{\sc ii} 
 lines to the 8 targets, we find that  two lines of sight situated close to 
 the nebula as mapped in CO and H\,{\sc i} exhibit very strong Na\,{\sc i} 
 absorption. This absorption is unlikely to arise in the foreground or background 
 interstellar medium which has a relatively  low column density, because the 
 Helix is at high galactic latitude $b = -57^{\circ}$. It is much more probable
 that it is due to the PN neutral or molecular material. This circumstellar origin 
 is reinforced by the fact that no corresponding Ca\,{\sc ii} line is observed 
 (as it would usually be from the ISM), which is in agreement with the very high 
 Ca depletion often observed  in PNs. No trace of any  DIB 
 features was found in these two circumstellar lines of sight,  nor to the other 
 targets. The two circumstellar sightlines discovered in this work open the way to search 
for molecular species such as C$_2$ which are not observable at radio wavelengths, 
and to obtain more information on the rich neutral and molecular content of the Helix.
   }
   \titlerunning{Optical absorption line spectroscopy of the Helix PN}

   \maketitle
%
%________________________________________________________________

\section{Introduction}

 The identification of the carriers of the diffuse interstellar bands (DIBs)
 has proved one of the most difficult problems in observational molecular 
 astrophysics (see Herbig \cite{her95}, hereafter H95, for a review).  
 DIBs have been widely 
 observed throughout the interstellar medium (ISM) for nearly a century, but 
 no convincing identification of a carrier has yet been forthcoming. In addition, 
 the origin of the carrier(s)  remains unknown.  Numerous studies have suggested 
 that DIB carriers are not present in the circumstellar matter of evolved cool 
 O-rich or C-rich stars that inject various dust or molecular compounds into the 
 ISM (Snow \cite{sno73}; Snow \& Wallerstein \cite{snw72}; Zacs et al. \cite{zac03}). 
 However, there are a few exceptions that would certainly  deserve further 
 attention, e.g. the  unusual 6177\,\AA~ band seen in some evolved objects by 
 Le Bertre \& Lequeux (\cite{leb93}), or the case 
 of DIBs possibly observed against the A-type companion of the carbon star CS\,776 
 (Le Bertre \cite{leb90}; for this object, see also comments in H95 and in Kendall et al. 
 \cite{ken02}).
 
The initial findings of Scarrott et al. (\cite{sca92}) and Sarre et al. (\cite{sar95}) 
connecting the optical emission 
bands in the Red Rectangle (RR) with a subset of well-known DIBs have been qualified 
recently by Van Winckel et al. (\cite{vwi02}) and Glinski \& Anderson (\cite{gli02}) on 
the basis of e.g. a 2\,\AA~wavelength 
mismatch between the $\lambda$5797 DIB in the ISM and the nearby $\sim$5800\,\AA\,\,
RR\,\,emission band. This mismatch remains at the largest observed distances from the central 
star, leading both authors to conclude that the  carriers of the DIB and the RR band 
are not identical. However, Van Winckel et al. (\cite{vwi02}) also note five  further possible 
matches between DIB wavelengths and RR features, and there certainly exist in 
the RR molecules closely related to the likely DIB carriers.

A hitherto undisputed fact is that a subset of diffuse bands are also clearly seen 
in emission in the circumstellar layers of the unusual, carbon rich, R\,CrB-type 
variable V854\,Cen (Rao \& Lambert \cite{rao93}). DIB carriers perhaps exist also 
in the envelopes of 
rare, hotter C-rich post-AGB objects (Zacs et al. \cite{zac99}; Klochkova et al.
 \cite{klo99}; Klochkova et al. \cite{klo00}; Klochkova et al.  
\cite{klo01}), although this is not certain.  A clear-cut case concerning 
a post-AGB object is HR\,4049,
which has a carbon-rich envelope; no diffuse band of circumstellar origin 
could be found in the spectrum of the central hot star (Waters et al. \cite{wat89}). 
Because it is often thought that large carbon bearing molecules might be the 
(or one of the) DIB carriers (e.g. H95, Duley \cite{dul98}), every effort is desirable to 
test  this hypothesis  by seeking DIBs at carbon-rich sites of origin
where these molecules may be abundant.

In a recent paper (Kendall et al. \cite{ken02}), we have pioneered a  new 
observational method using optical absorption spectroscopy of background objects to 
probe the spatially extended circumstellar envelope (CSE) of the archetypal 
extreme mass-losing carbon star IRC\,+10$^{\circ}$\,216. By searching for absorptions 
against the relatively smooth continua of thick-disk background targets, the method 
alleviated confusion with photospheric features in the highly complex spectrum of 
the central, cool, evolved object. These observations showed that  DIB carriers are 
of very low abundance, or non-existent, around IRC\,+10$^{\circ}$\,216, and we 
suggested that the DIB carriers might be sought in compact, UV-rich regions where ionised 
and neutral molecular material interact. Such sites are the circumstellar shells of 
planetary nebulae (PN), and this paper reports our first attempts to detect DIBs in 
the Helix PN. 

In fact, DIB carriers have already been sought in a few other carbon-rich PN, most 
notably NGC\,7027~(Le Bertre \& Lequeux \cite{leb92}) and no circumstellar diffuse 
bands  have been detected.  However, it is justified to apply our method of background 
target spectroscopy to PN and especially to use several lines of sight.
 This is an important point, since the clear clumpiness and inhomogeneity
of typical PN shells means that DIB carriers, and other species,  may not be distributed 
uniformly throughout such shells. Indeed, in cases where only
the central star is observed  through relatively low density material, the column density 
may be insufficient  for observable absorptions to arise. 

Therefore,  in this paper, we report  the first results of a
similar absorption spectroscopy  experiment performed using 8 targets seen through, or
located angularly very close to, the Helix. As explained below in more detail (Sect. 2), this  
PN was selected primarily because of its proximity, its high galactic 
latitude, and its wealth in ionized, neutral and molecular content.

%%%==================================== TABLE OF TARGETS ===================================
\begin{table*}
 \caption[]{Background targets beyond the Helix Nebula, with basic and derived data. 
  For clarity, we will use the identification given in the second column when referring 
  to individual targets.  
The column t$_{\rm exp}$ gives for each target
the total exposure time; for objects 1, 3, 4, and 8, two spectra of 
half this time were exposed. $d$ is the estimated distance in pc.
The final column states whether  a target is directly behind the nebula (on-PN) 
or a control object (off-PN). Targets 7 and 8 are   HD\,213056 and HD\,213069, 
respectively, with spectral types  from Simbad.}
\begin{center}

        \begin{tabular}{crccrrrrrcr}
        \noalign{\smallskip}
        \hline
        \hline
        \noalign{\smallskip}
IAU  &Id. & $\alpha$\,\,  $\delta$ & $V$ & $B-V$ & $V-J$  & t$_{\rm exp}$ &Sp.& $d$ &
 Notes \\
name & & (J2000)             & (mag)&(mag)&(mag)    & (min)          &      & (pc)   &\\
 
\noalign{\smallskip}
\hline
\noalign{\smallskip}
USNO-B1.0 0691-0897349& 1 & 22 29 47.3 $-$20 50 31 & 13.17 & 0.51  & 1.02 & 66.2 & F8V & 680 &  on PN\\
USNO-B1.0 0691-0897340& 2 & 22 29 38.6 $-$20 50 14 & 13.54 &$-$0.39& -\,\,& 46.1 &-\,\,& 210 & nucleus\\
USNO-B1.0 0692-0882100& 3 & 22 29 33.5 $-$20 46 18 & 12.04 & 0.64  & 1.08 & 22.7 & G2V & 290 & on-PN\\
USNO-B1.0 0691-0897309& 4 & 22 29 19.0 $-$20 48 48 & 13.94 & -\,\, & 0.75 & 66.7 & F2V & 1500& on-PN\\
USNO-B1.0 0690-0897407& 5 & 22 29 46.5 $-$20 56 33 & 13.57 & -\,\, & 0.68 & 47.4 & F0V & 1500& on-PN\\
USNO-B1.0 0691-0897499& 6 & 22 30 28.0 $-$20 48 31 & 12.31 & 0.59  & 1.16 & 14.6 & G1V & 380 & off-PN\\
USNO-B1.0 0692-0882044& 7 & 22 29 09.4 $-$20 46 07 & 10.02 & 0.89  & 1.88 & 01.7 &K1III& 760 & on-PN\\
USNO-B1.0 0690-0897288& 8 & 22 29 13.1 $-$20 57 49 & 10.36 & 0.73  & -\,\,& 05.2 & F5V & 240 & off-PN\\
\noalign{\smallskip}
\hline
\end{tabular}
\end{center}
\end{table*}
%%%%%%%%%%%%%%%%%++=======================================================================================

%=========================== figure showing Helix; graph bidon pour ArXiv
   \begin{figure}
   \centering
   \resizebox{\hsize}{!}{
  {\rotatebox{-90} {\includegraphics{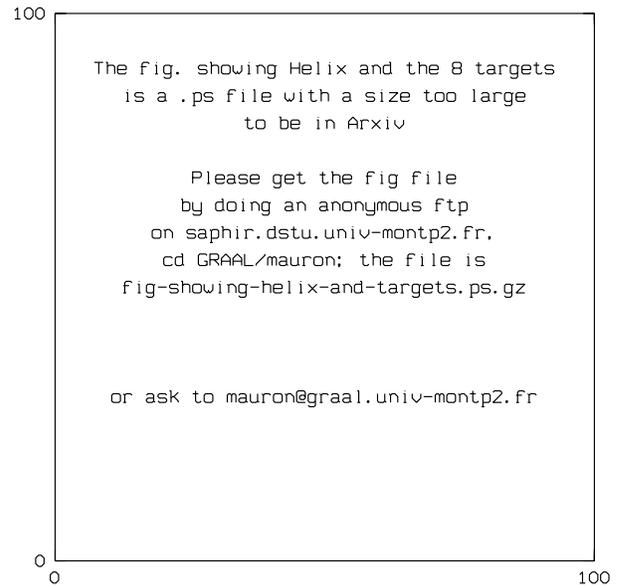}}   }
  } 
  \caption{Observed targets labelled on the $R$-band POSS-2 image. 
        The field  is 25$^{\prime}$ $\times$ 25$^{\prime}$, centered on
       the nucleus (\#2). North at top, East to the left.}
   \label{f1}
   \end{figure}
%%%%%%%%%%%%%%%%%%%%%====================================================================

\section{The Helix Nebula and the background stars used for absorption spectroscopy}

The Helix Nebula is undoubtedly the best target of its kind for this study, being 
close ($d$\,=\,$210\pm30$\,pc, from parallax $4.70\pm0.75$\,mas; Young et al. \cite{you99}), 
as well as rich in carbon, dust and being irradiated by a strong central UV source. 
Its high galactic latitude ($b=-57^{\circ}$)
minimises confusion with the foreground or background ISM: the maps of 
Burnstein \& Heiles (\cite{bur82}) and Schlegel et al.  (\cite{sch98})
 show a very low interstellar (IS) reddening of
$E_{\rm B-V}$\,$\sim$\,0.03\,mag. The nebula has been detected in the lines of 
H\,{\sc i}, H$_2$ and CO, together with other species (CN, HCN, HNC, HCO$^+$
and C\,{\sc i}) which strongly indicate a carbon-rich medium (Young et al. \cite{you99}; 
Speck et al. \cite{spe02}). 
While the H\,{\sc i} and CO\,(2--1) maps (Rodriguez et al. \cite{rod02}; 
Young et al. \cite{you99}, respectively) 
roughly outline the main body of the optical PN, i.e. as traced by ionised species,
these maps reveal strong inhomogeneity in the form of a marked ``knotty'' structure. 
This is fully consistent with the presence of the well known numerous 
dust globules seen in optical images. Also, the radio maps show that the global PN 
geometry is not spherical, but approximately disk-like and seen nearly pole-on. This 
complex structure makes obvious the necessity of considering as many  background 
stars as possible. Finally, because the neutral phase of the PN   contains $\sim$\,3 
to 6 times more mass than the molecular phase (Rodriguez et al. \cite{rod02};
 Young et al. \cite{you99}), it can 
be expected that essentially the neutral material, but perhaps also some molecular 
knots, may be probed by observing  background stars.

Optical images (Fig.\,\ref{f1}) show a large  number of potential targets within or 
near the PN shell which might function as sightlines to probe the nebula and the ISM 
in angularly nearby sightlines. Targets were first selected using homogeneous, accurate
$JHK$ photometry from 2MASS. 
For some of them, a $V$ magnitude is available in Simbad.
By comparing the $J$$-$$K$ and/or $V$$-$$J$ colours of our targets to 
template Population\,I dwarfs, an approximate spectral type could be inferred. 
Then, absolute magnitudes and therefore distances were derived. Where independent 
$U$$-$$B$, $B$$-$$V$ colours are available in Simbad, we find excellent agreement with 
the predictions of the $J$$-$$K$ or $V$$-$$J$ colours. 
Only about half of the targets which were 
bright enough to obtain high signal-to-noise ($S/N$) spectra could eventually be observed, 
owing to poor weather.
 
The derived types and distances for these 8 observed targets are given in Table\,1. 
The targets  have $V$\,$\sim$\,10--14 and are located directly behind, or 
close to, the shell.  With the exception of the nucleus, target \#2, 
the sightlines probed have spectral types F or G. The  uncertainties on distances are
of the order of 20\%. Distances would be 30\% smaller (or less) if we had adopted 
the luminosity of low-metallicity subdwarfs, but they would be considerably larger
if we had adopted the luminosity of giants. In the end, several targets are clearly much more 
distant ($>$\,400\,pc) than the Helix. As we shall see below, all  these targets  provide
a very usable continuum for absorption spectroscopy 
despite  their photospheric lines.  Moreover, for each target,
 the precise velocity of the photospheric Na\,{\sc i} D lines could be easily predicted
through consideration of  
$\sim$20 prominent unblended  Fe\,{\sc i} lines available in our spectra 
(vertical bars in Fig.\,\ref{f2}).

\section{Observations and data reduction}

Observations were performed on 2002 October 11--12 using the Very Large 
Telescope with the Ultraviolet and Visual Echelle 
Spectrograph (UVES) at the European Southern Observatory, Cerro Paranal, Chile. 
Total integration times of up to $\sim$\,one hour were achieved, depending on the 
target magnitude. Using the 0.8\arcsec~slit (R\,=\,50000) with both red 
and blue dichroics, spectra were obtained over three wavelength ranges, 
$\sim$\,3500--4520\,\AA, 4630--5600\,\AA~ and 5670--6650\,\AA.
These ranges include in particular the strongest DIBs at 5780, 5797, 6283 
and 6614\,\AA,  the Na\,{\sc i} D and Ca\,{\sc ii} H \& K lines. 
The target signal-to-noise ratio of $\sim$\,70 (per 0.017\AA\,bin) was achieved 
over the most part of these ranges, except in the blue region
comprising the Ca\,{\sc ii} lines where $S/N$ is generally poor because
the target is not hot enough (with the exception of the nucleus).
The data analysis was carried out by using the
reduced spectra  that are provided by the ESO/UVES reduction pipeline, and by
using the ESO MIDAS and the NOAO IRAF packages.
 
%%%%%%%%%%%%%%%%%%%  TABLE ON 3 FIELD HOT STARS  =================================
\begin{table*}
 \caption[]{Literature data for 3 bright stars located at 2.8--5.4\degr\, 
            from the Helix. $\theta$ is the angular distance of each star
	    to the Helix nucleus. W$_{\lambda}$(D1) \& W$_{\lambda}$(D2) are the observed 
	    equivalent widths in the IS Na\,{\sc i} D1 5895\,\AA\, and D2 5890\,\AA\, 
	    lines, and W$_{\lambda}$(K) is that for the Ca\,{\sc ii} K 3933\,\AA\, line, all 
	    expressed in m\AA. Relative uncertainties on 
	    W$_{\lambda}$ values are $\sim$ a few percent.}
 \begin{center}
 \begin{tabular}{lcrlrrrrl}
 \noalign{\smallskip}
 \hline
 \hline
 \noalign{\smallskip}

 Star      & $\theta$ &  V\,\,\,\,\,   & Sp.&dist. &W$_{\lambda}$(D1)& W$_{\lambda}$(D2)&
 W$_{\lambda}$(K) & Notes\\
 name          &  (deg)   & (mag)&       & (pc) &(mA)  & (mA)  & (mA)     & \\ 
\noalign{\smallskip}	     
 \hline
 \noalign{\smallskip}
 PHL\,346   & 2.86     &11.47 & B1    &10000 &  -\,\,   &    -\,\,  &  199     & (1)(4)\\
 
 HD\,214080 & 4.70     & 6.83 & B1Ib  & 3000 & 171      &   207     &  149     &  (2)\\
 
 HD\,210191 & 5.38     & 5.81 & B2.5IV&  460 &  90      &  164      &  285     & (2)(3)\\
\noalign{\smallskip}
\hline
\end{tabular}
\end{center}

Notes: (1) Data from Keenan et al. (\cite{kee88});    (2) Data from Albert (\cite{alb83});
 (3) Also known as 35 Aqr; (4)  for PHL\,346, no Na\,{\sc i} data could be found 
 in the literature.
 \end{table*}
%%%====================================================================================

\section{The interstellar medium in the direction of the Helix Nebula}

  Before interpreting the spectra of our program stars, e.g. 
 the Na\,{\sc i} and Ca\,{\sc ii} profiles, 
 one has to estimate the effect of the  foreground and background 
 ISM, deriving, at least approximately, the expected strength of the IS lines.
 One  source of information is the $E_{\rm B-V}$ maps derived 
 by Schlegel et al. (\cite{sch98}).  By considering a zone of
 3\degr$\times$\,3\degr~centred on  the Helix (the nucleus is at  $l$\,=\,36\fdg2,
 $b$\,=\,$-57$\fdg1)   but excluding a central region of 
 1\degr$\times$\,1\degr~to avoid the nebula itself,
 we find that $E_{\rm B-V}$ would be on average $\approx$ 0.035 mag. 
 with $\sigma$\,=\,0.005, as derived  from  20 independent positions.
 One has to note that the $\sigma$ above only represents the local 
 scatter of $E_{\rm B-V}$ as given by the Schlegel et al. maps which have an
 angular resolution of 6\arcmin. This does not exclude  
 moderately larger deviations from this mean when one considers the very narrow 
 pencil-like sightlines to  background stars. 
 
 Based on this estimate of $E_{\rm B-V}$, one can derive 
 the expected column density of Na\,{\sc i}. From Bohlin et al. (\cite{boh78}), 
 the hydrogen (H\,{\sc i} + 2H$_2$) column density is obtainable with  
 $N$(H)/$E_{\rm B-V}$ = 5.8\,$\times$\,10$^{+21}$ cm$^{-2}$, with
  deviations from this mean generally less than a factor 1.5. Consequently, 
  we obtain $N$(H)\,=\,2.0\,$\times$\,10$^{+20}$ cm$^{-2}$. For this column 
 density, the correlation found by Ferlet et al. (\cite{fer85}) with $N$(Na\,{\sc i})
 suggest that $N$(Na\,{\sc i})\,=\,1.0\,$\times$\,10$^{+12}$  cm$^{-2}$ with  
 a scatter of about a factor 3 (see their Fig.\,1). This Na\,{\sc i} column density 
 is relatively small but is already in the non linear regime of the curve of 
 growth for typical IS sightlines.  It corresponds to a typical 
 equivalent width (W$_{\lambda}$) in the Na\,{\sc i} D2 (5890\,\AA) line 
 W$_{\lambda}$(D2)\,$\sim$\,200\,m\AA. Consequently, we expect that 
 W$_{\lambda}$(D2) observed towards our targets  be close to 200\,m\AA.

 Another source of information  concerning the ISM in the direction of the Helix 
 is obtained by considering available observations of Na\,{\sc i} or other  
 atoms for stars angularly close to the Helix. By interrogating the catalogue of 
 Garcia (\cite{gar91}), we find 3 stars with IS information available 
 and located  at less than 6\degr~from the Helix centre. The data, taken 
 from Albert (\cite{alb83}) and Keenan et al. (\cite{kee88}), are listed in Table 2, and
 the values of W$_{\lambda}$(D2) are found to be also  $\sim$\,200\,m\AA\, in
 excellent agreement with estimations made above. It is interesting also to note 
 that Albert (\cite{alb83}) gives for HD\,214080 $E_{\rm B-V}$\,=\,0.05  in very 
 good agreement with $E_{\rm B-V}$\,=\,0.047 provided by the Schegel et al. maps 
 in the direction of this star. In contrast, for HD\,210191, Albert (\cite{alb83}) gives 
 $E_{\rm B-V}$\,=\,0.07 whereas the maps of Schlegel et al. give 
 $E_{\rm B-V}$\,=\,0.025. It is possible that Albert (\cite{alb83}) overestimates the
extinction because the intrinsic colour $(B-V)_{0}$ of HD\,210191 has 
to be derived  from its spectral classification and may have been 
incorrectly taken as too blue. 

An equally important point to consider is the expected range of heliocentric 
velocities for IS lines. By considering Fig.\,2 of Albert (\cite{alb83}), it is clear that
the H\,{\sc i} (21\,cm) line, the Na\,{\sc i} and the Ca\,{\sc ii} lines toward 
HD\,214080 and HD\,210191 are all consistently 
found within the velocity interval $v_{\rm  LSR}$\,=\,$-$25 km\,s$^{-1}$ 
to +20\,km\,s$^{-1}$,
which corresponds to $v_{\rm helio}$\,=\,$-$28 to +17\,km\,s$^{-1}$
(for the direction of the Helix $v_{\rm helio}$\,=\,$v_{\rm  LSR}$\,--3, in km\,s$^{-1}$).
Concerning PHL\,346, Keenan et al. (\cite{kee88}) find the Ca\,{\sc ii} K IS line 
at $v_{\rm LSR}$\,=\,+21\,km\,s$^{-1}$, or $v_{\rm helio}$\,=\,+18\,km\,s$^{-1}$, 
in good agreement with the other two bright stars. 

To conclude, the above considerations  suggest that IS lines in our target 
sightlines should display typical W$_{\lambda}$ of 200\,m\AA~ for Na\,{\sc i} D2, and
about the same for the Ca\,{\sc ii} K line (see Table 2). We expect
the IS lines to have $v_{\rm helio}$ within the interval 
from $-$28 to +17\,km\,s$^{-1}$, although a velocity slightly outside of this expected
range is not excluded, because it is based on 3 sightlines which
are not exactly toward the Helix, but angularly shifted by a few degrees.

%%%%%%%%%%%%%%%%%%%%%  figure 2  showing NAI and Ca\,{\sc ii} profiles ===========
\begin{figure*}

\resizebox{\hsize}{!}{
{\rotatebox{-90}{\includegraphics{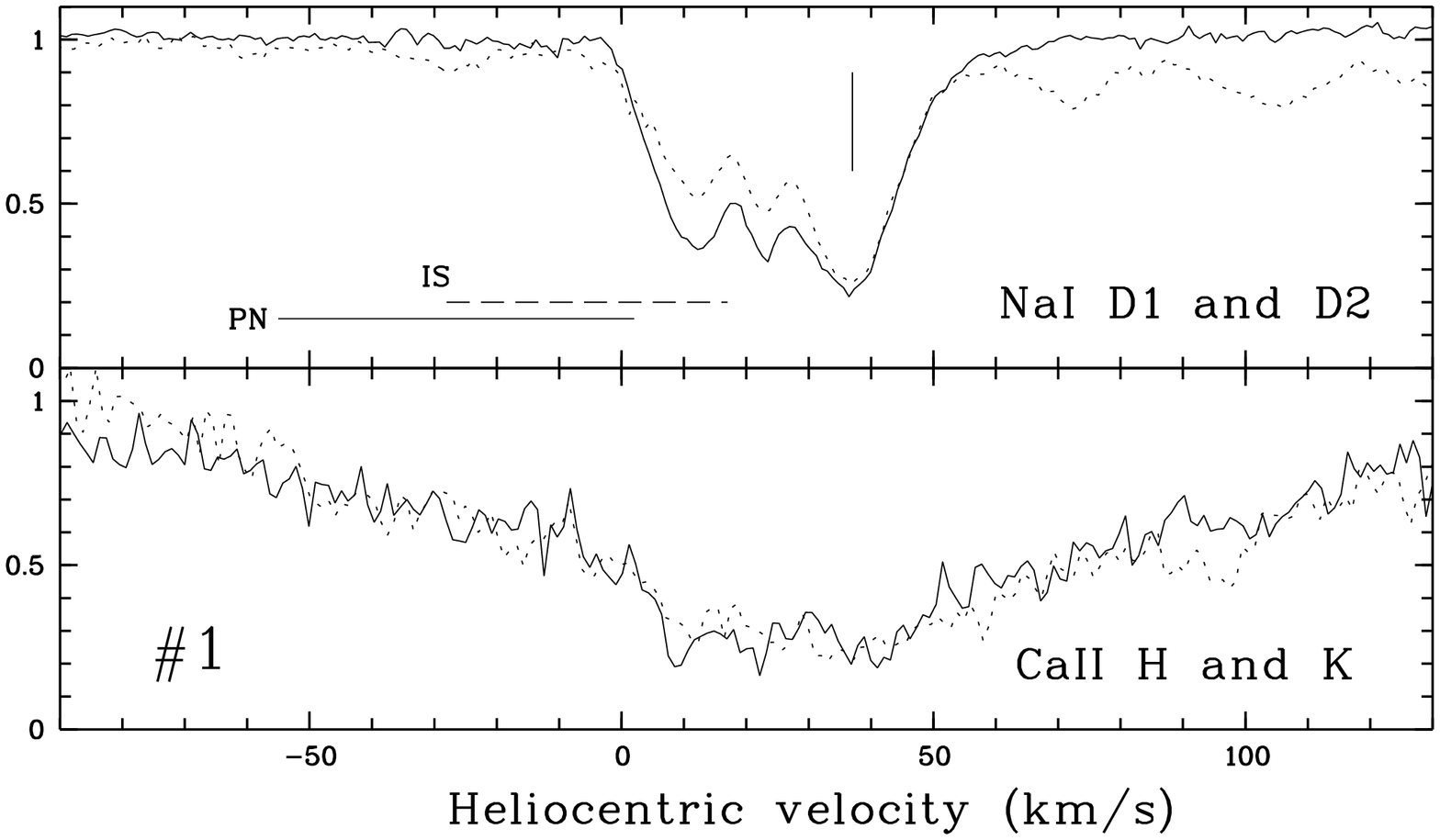}}}
{\rotatebox{-90}{\includegraphics{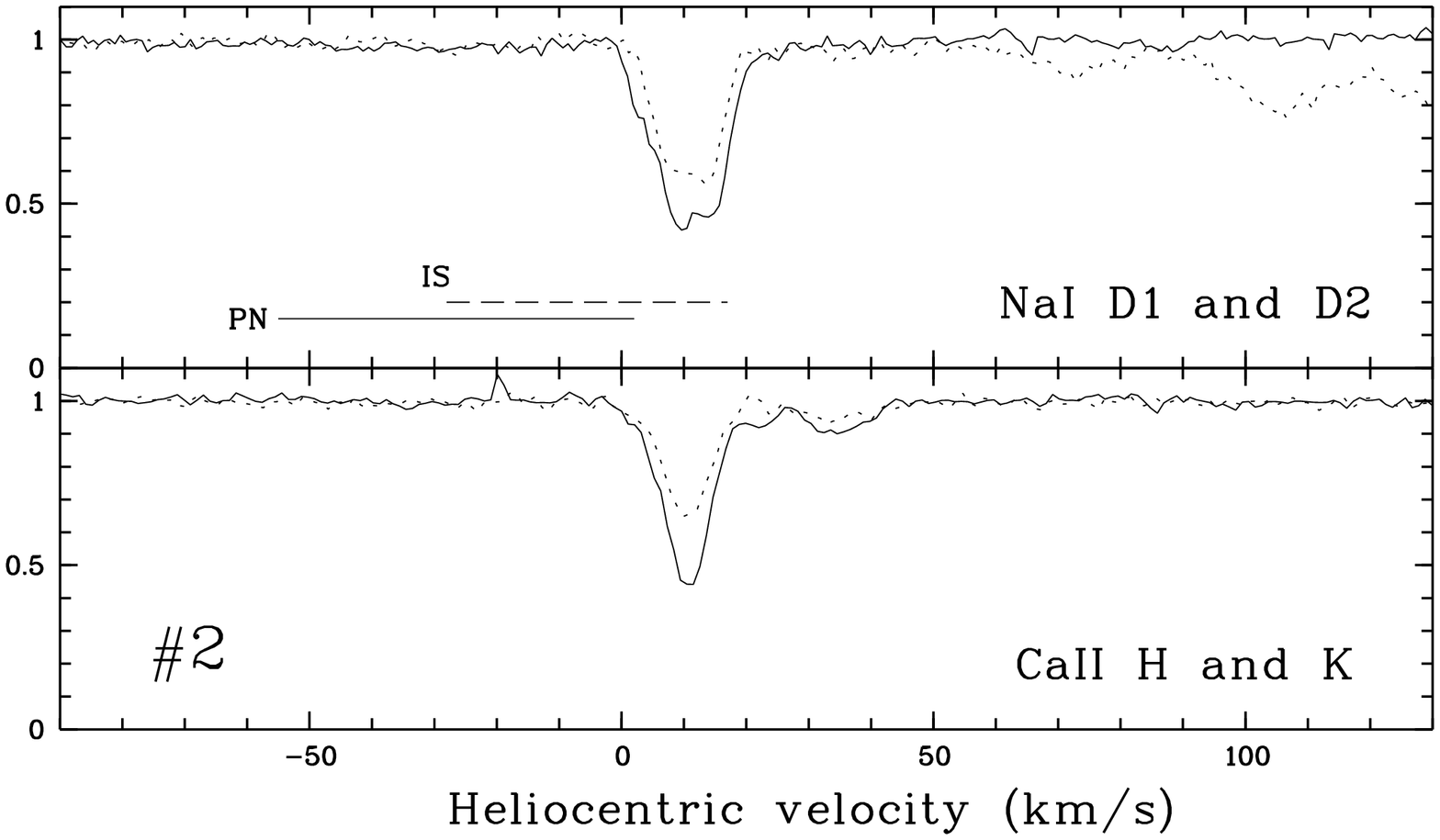}}} }

\resizebox{\hsize}{!}{
{\rotatebox{-90}{\includegraphics{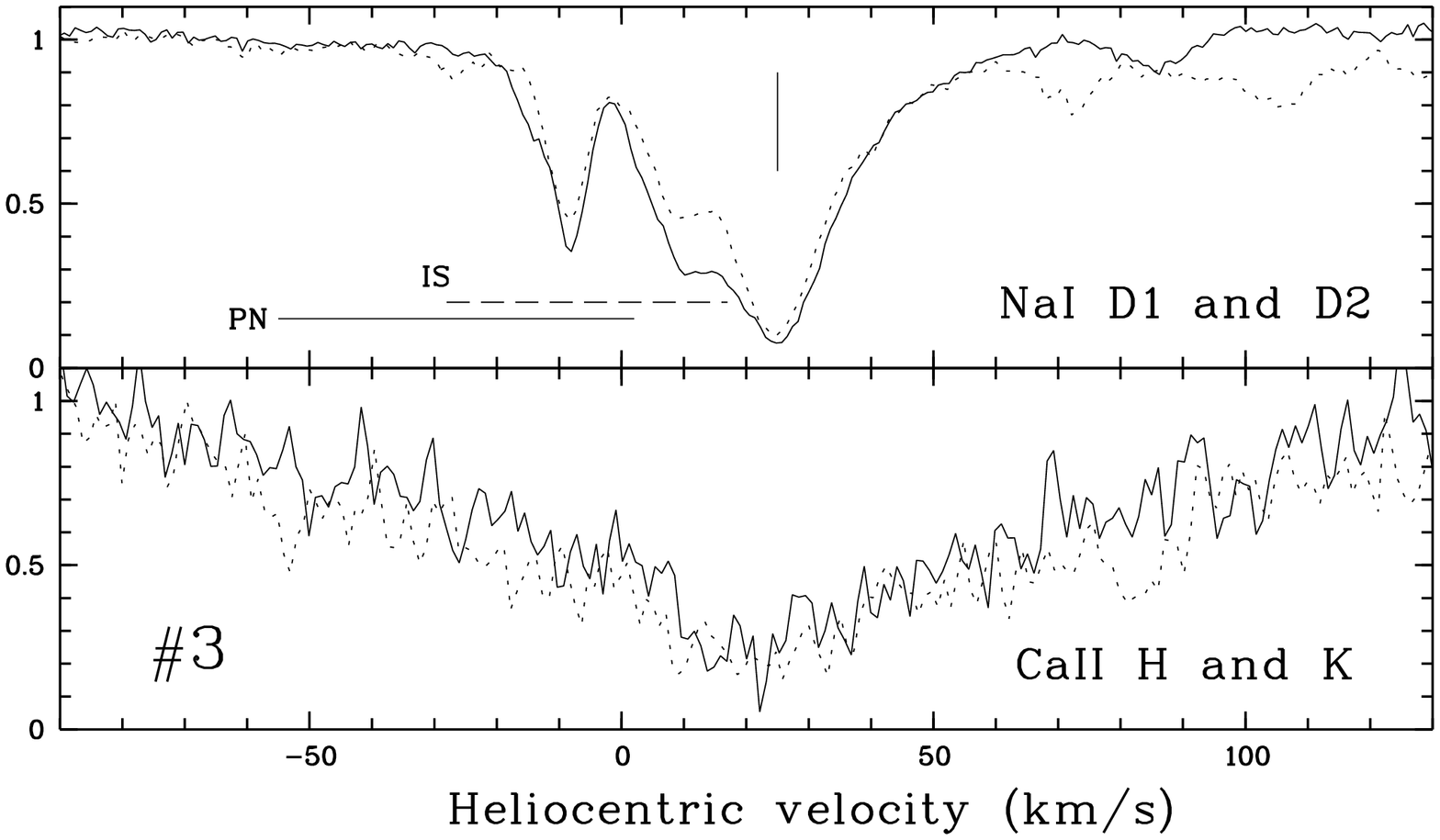}}}
{\rotatebox{-90}{\includegraphics{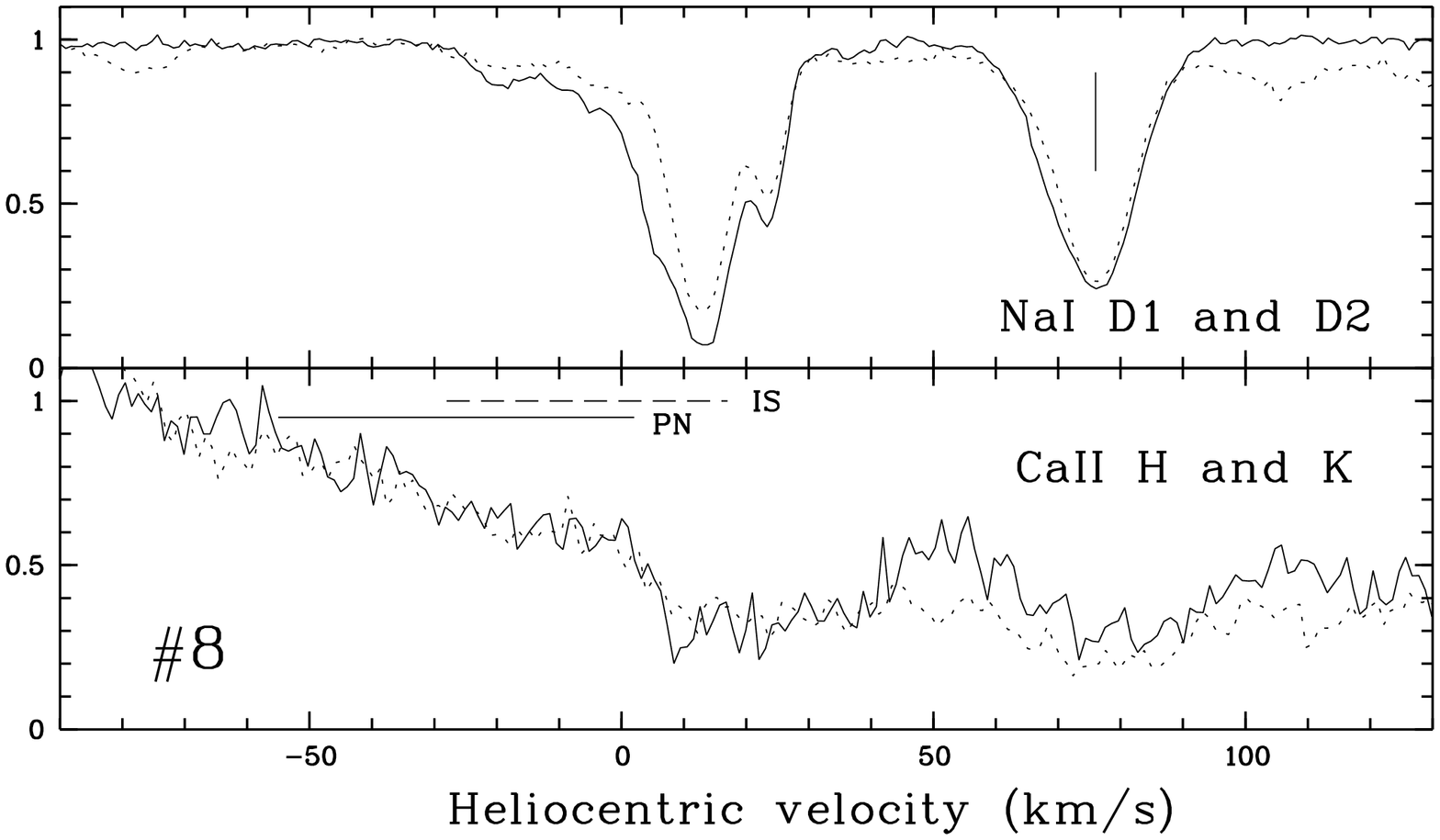}}} }

\resizebox{\hsize}{!}{
{\rotatebox{-90}{\includegraphics{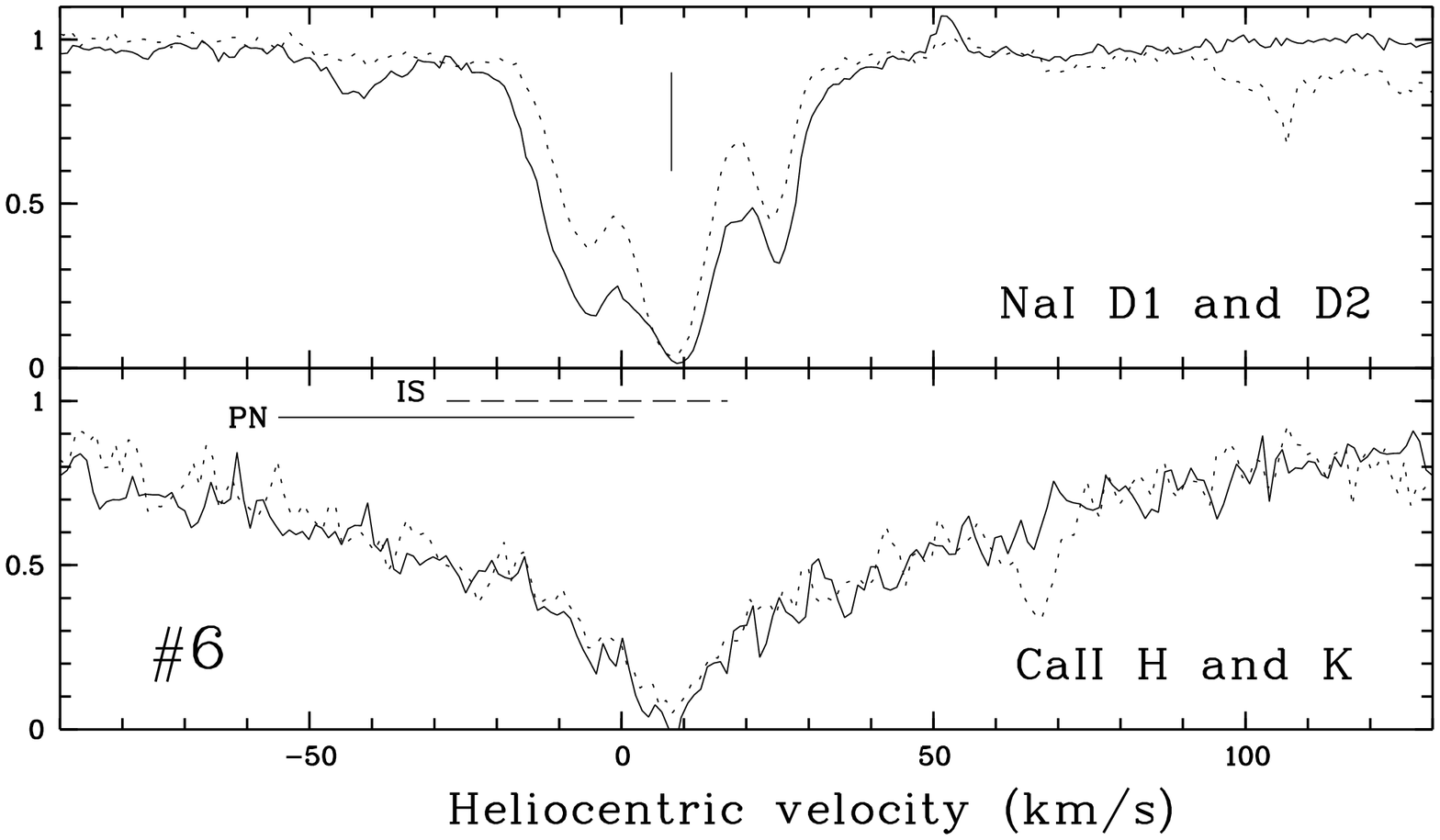}}}
{\rotatebox{-90}{\includegraphics{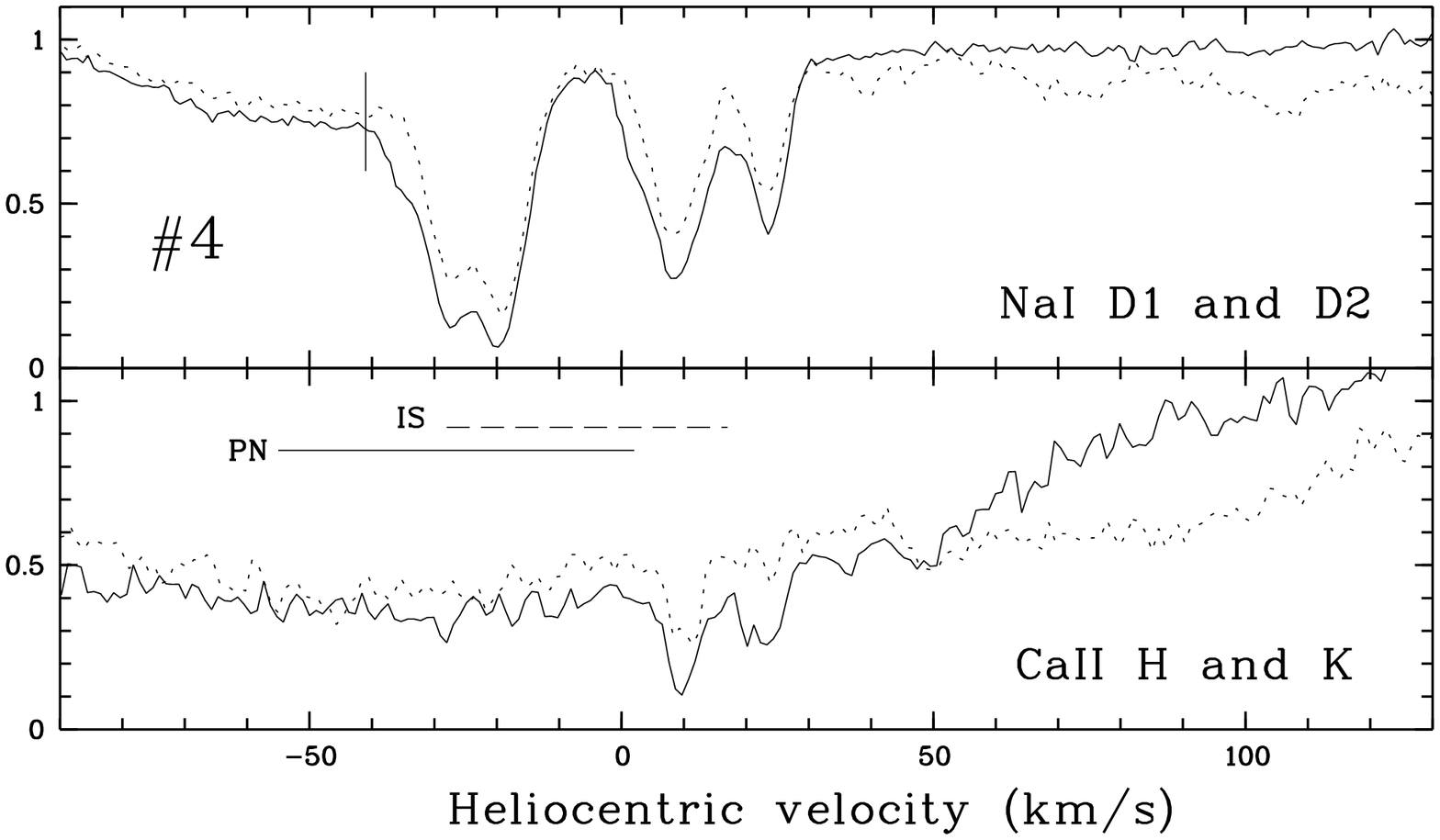}}} }

\resizebox{\hsize}{!}{
{\rotatebox{-90}{\includegraphics{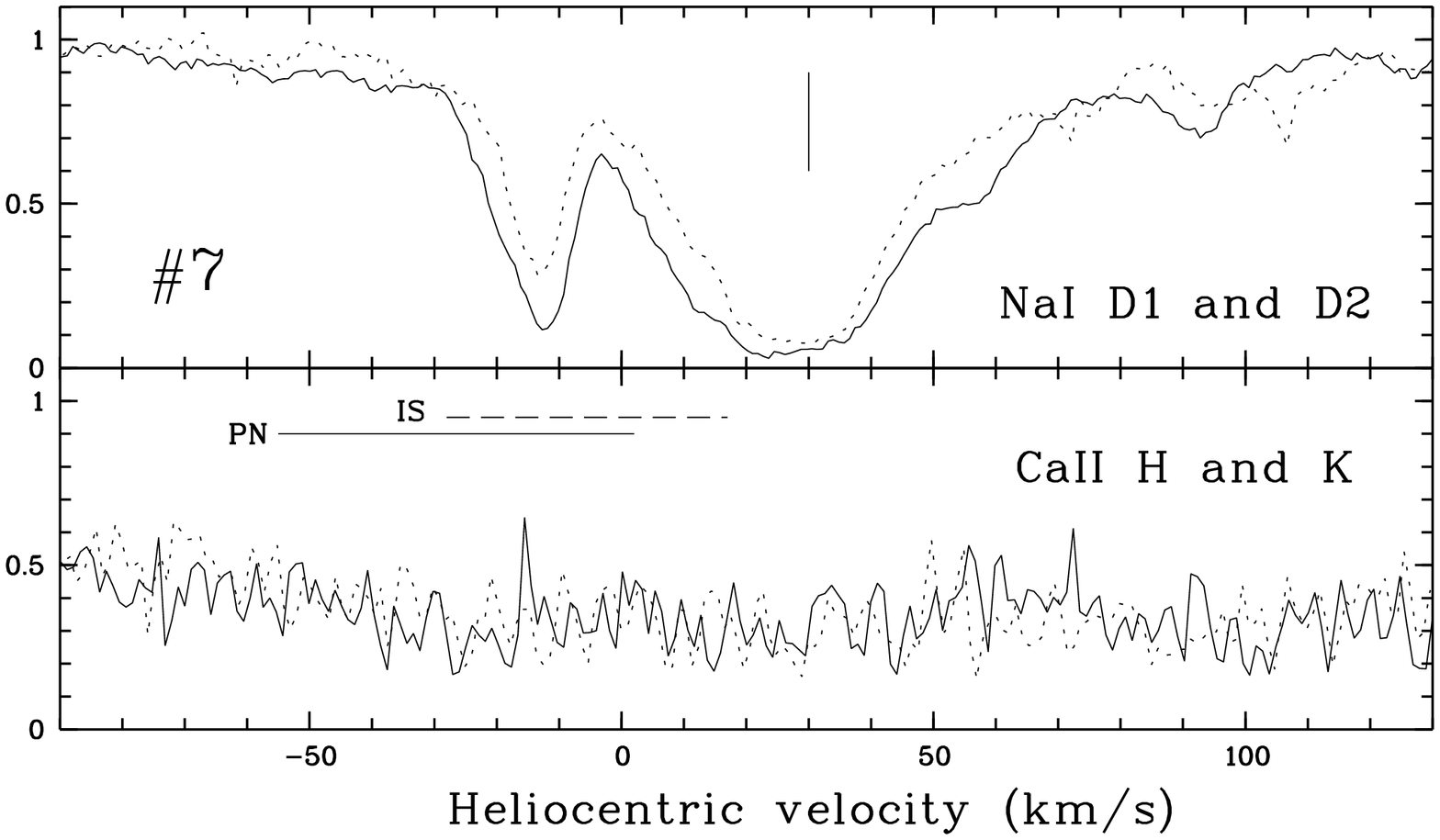}}}
{\rotatebox{-90}{\includegraphics{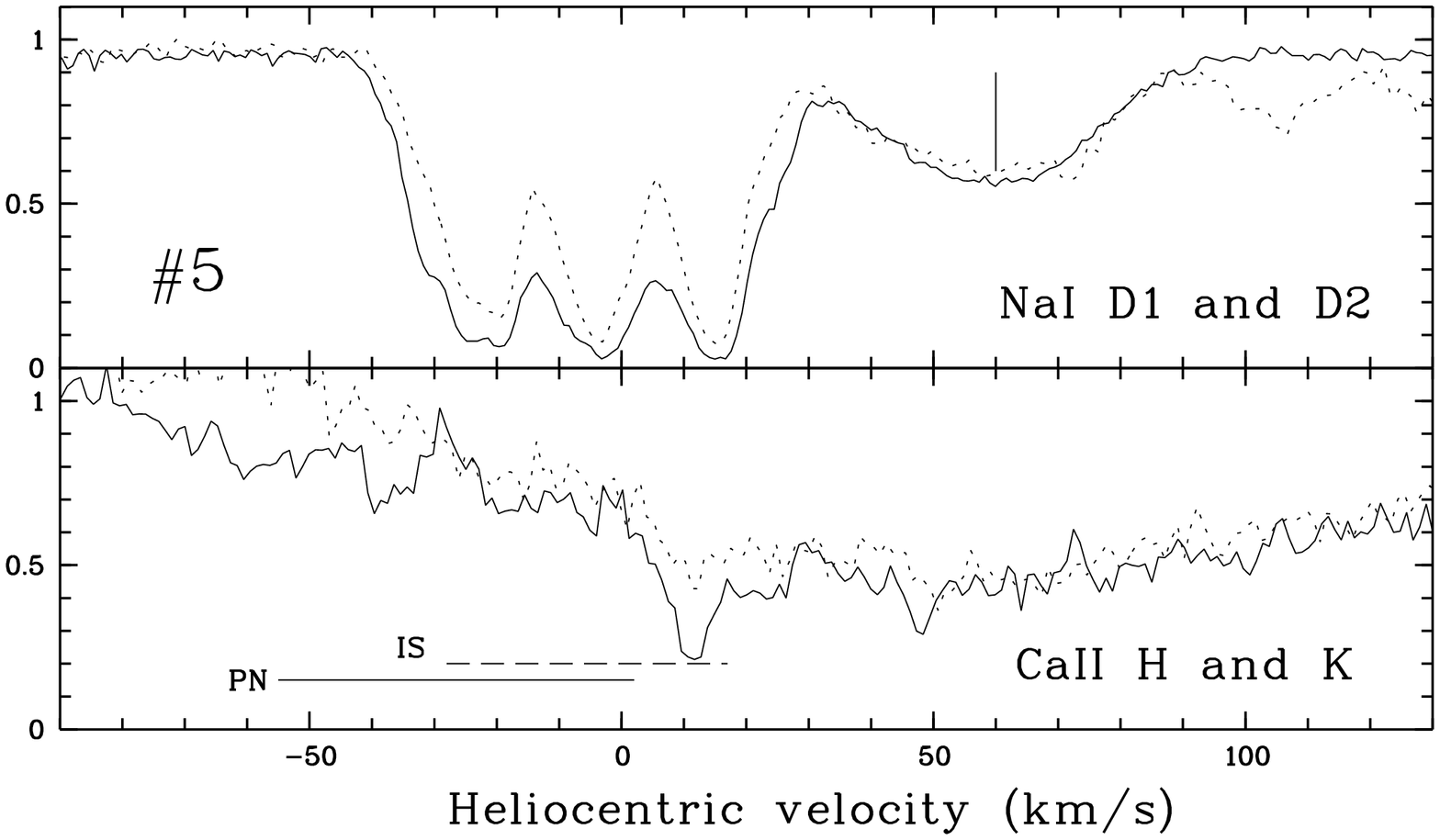}}} }

\caption[]{Na\,{\sc i} D1 (5896\AA) and D2 (5890\AA) profiles (upper plots) 
and Ca\,{\sc ii} H (3968\AA) and K (3933\AA) profiles for all eight targets. 
For Na, D2 is the solid line and D1 the dotted. For Ca, the solid line 
is K, the dotted H. The velocity of photospheric Na \& Ca lines is 
shown by a vertical bar. In the left-hand four panels, we show targets \#1, 
\#3, \#6 and \#7 for which we suspect there is no circumstellar Na D component. 
The same is true of the nucleus (\#2, upper right) where we see 
a pure IS spectrum because of the lack of photospheric contamination. 
For \#8, the photosphere is well separated, at $\sim$\,+76\,km\,s$^{-1}$. 
For \#4, the component at $-$25\,km\,s$^{-1}$ is most likely nebular, as 
is most of the Na\,{\sc i} line for \#5 (see text).} 
\label{f2}
\end{figure*}
%%%%==============================================================================

\section{Results and analysis}

\subsection{Characterising the nebular absorption using Na\,{\sc i} D profiles}

Figure\,\ref{f2}
displays the Na\,{\sc i} D1 and D2 line profiles for all eight targets.
It is important to realize first that, for a given target, any detail seen in 
{\it both} D1 and D2 profiles should be the signature of the same IS (or possibly
circumstellar) component, and one can indeed see in these spectra many
components with corresponding fully consistent D1 and D2 features. 
This compensates for one of the difficulties of our experiment:
one has  to be  careful about  the fact that the stellar continua of our targets 
are not flat, as they would be for OB type stars. Consequently,
depending on the radial velocity of the targets, the Na\,{\sc i} 
IS (or circumstellar) lines may
be blended with photospheric Na\,{\sc i} lines in the target stars themselves, 
or, possibly, with a different photospheric feature. 
Finally, since Na\,{\sc i} has now been detected in a number of PNs 
(Dinerstein \& Sneden \cite{din88}; Dinerstein et al. \cite{din95}), 
one can expect some circumstellar contribution, that could well be 
significant. 

 In Fig.\,\ref{f2}, the approximate range of velocities for IS components 
(discussed above) is indicated by a horizontal dashed line labelled ``IS''.
 Concerning the velocities of possible circumstellar components, we have
 plotted with a horizontal solid line labelled ``PN'' the $v_{\rm helio}$ range 
 of the Helix gas: this range is  from   $-$55 to +2\,km\,s$^{-1}$, as derived 
 from CO and H\,{\sc i} line profiles (integrated over the PN). 
 In principle, in order to predict at which 
 velocity one could expect a circumstellar component, one might consider the 
 CO or H\,{\sc i} maps and look at the profiles at the exact position of each 
 target. In practice, this provides little information: none of the 8 targets are
 covered by the H\,{\sc i} map of Rodriguez et al. (\cite{rod02}; their Fig.~4), 
 and CO is detected only to  target \#3, after a careful examination of  Fig.~1 of 
 Young et al. (\cite{you99});
 the case of target \#3 is discussed below. Therefore, we adopt the view that
 circumstellar Na\,{\sc i} components may  occur anywhere within the CO or H\,{\sc i} 
 velocity range.

 \subsubsection{Targets 1, 3, 6 and 7 (Fig.\ref{f2}, left panels)}

The case of blending with photospheric Na is obviously present
for targets \#1, \#3, \#6 and \#7. In the light of the
above comments, these four targets
show nothing exceptional which is not explained by photospheric
{\it plus} IS contributions. 
We note one component at $v_{\rm helio} \approx +25$\,km\,s$^{-1}$ 
for  \#1 and \#6. This is +7\,km\,s$^{-1}$ outside 
 the expected velocity range of IS lines, and not near the $v_{\rm helio}$ 
 of the Helix itself or that of the target photospheres. It may be a local 
 feature of the ISM affecting only these lines of sight. 

There is also  a distinct component at 
$v_{\rm helio}$\,$\approx$\, $-10$\,km\,s$^{-1}$ 
 for \#3 and \#7, and this might {\it a priori} be assigned to
 IS matter in these lines of sight, since their W$_{\lambda}$(D2) are
 $\sim$\,100 and 230\,m\AA~for \#3 and \#7 respectively, and because
 this velocity falls in the expected range for IS absorptions. The relatively small 
value for \#3 (suggesting a lack of a circumstellar contribution) 
is surprising in view of the location of this target within the CO isophotes
(see Fig.~1 of Young et al. \cite{you99}).This line of sight should trace 
neutral/molecular nebular material, but this is not apparent in Na\,{\sc i} and 
it may be that the Helix is located beyond this star, which is the nearest of 
all the targets. 
In any case, the component is not at the observed CO velocity
in this region, between $-$30 and $-$50 \,km\,s$^{-1}$, see Fig.~3 of 
Young et al. (\cite{you99}). 
For target \#7, well off the PN and CO contours, the strength
of this particular component is typical of the ISM in the direction of the Helix.

\subsubsection{ Targets 2, 8 (Fig.\,\ref{f2}, upper right panels)}

  One can pursue this analysis further by considering target \#2 (the PN nucleus)
 and target \#8, for which no blending with photospheric lines
 exists. For the nucleus, this is because no photospheric Na\,{\sc i} line is expected, 
 given its likely very hot $T_{\rm eff}$. Target \#8 is cooler with an 
 estimated F5V spectral type, but its  $v_{\rm helio}$ is such 
 (+76\,km\,s$^{-1}$) that the Na\,{\sc i} components are unaffected by any 
 photospheric Na line, or indeed any other line, since the D1 and D2 
 profiles resemble each other accurately.
 The distances of these two stars are  $\sim$\,210--240 pc.
 For \#2, W$_{\lambda}$(D2) is found to be  $\sim$\,150\,m\AA, and we
  do not expect any circumstellar absorption, because of the PN geometry, 
or photospheric absorption, so by definition this is the ISM probed towards the nucleus 
itself, at its fairly well-known distance. 

For target \#8, well off the PN, the
photospheric and (presumably) IS components are well separated (as mentioned above) 
and we can measure the  photospheric (D2) strength as $\sim$\,250\,m\AA~(seemingly
typical of all our F--G type targets). The IS component is surprisingly strong at 
410\,m\AA. Because the absorption velocities in both \#2 and \#8 are  
at $\sim$\,+15 km\,s$^{-1}$, not within the Helix $v_{\rm helio}$ range, 
the simplest  interpretation is that we are again tracing  the foreground ISM, although
one cannot entirely rule out a nebular contribution.

\subsubsection{ Targets 4 and 5 (Fig.\,\ref{f2}, lower right panels)}
 
 Finally, targets \#4 and \#5 are particularly interesting for our 
 purposes because they are distant ($\sim$\,1500\,pc), and because they 
 are located close to the edge or angularly slightly 
 outside the bulk of the ionized PN gas. They are close to 
 the lowest contours of the available H\,{\sc i} and CO maps. Therefore,
 one can expect that these line of sight probe some neutral PN gas. Examination
 of the data indeed suggest this is the case.
 
Given the strength of the Na\,{\sc i} absorptions towards \#5, and that 
their velocity range  lie largely within the PN CO velocity range, we are led to
 attribute at least a significant part of the absorption  to circumstellar
material. The case of \#4 is perhaps even more clear cut; there is a strong, 
well-resolved component centred very closely
on the $v_{\rm helio}$ of the Helix  which lies at the (negative) velocity limit 
of where IS absorptions are expected. This component is at 
$\sim$$-$25\,km\,s$^{-1}$; the equivalent width of the individual component 
in D2 is 350\,m\AA, exceeding the canonical ISM expectation by nearly a factor of 2. 
Similarly, we note that \#5 has by far the greatest total Na\,{\sc i} equivalent 
width,  in both D1 and D2, of all the targets, certainly where the photosphere 
is well removed in velocity (\#7 has comparable absorption strength
but includes a significant photospheric component). This in itself lends support to 
the thesis that we are probing nebular Na\,{\sc i} towards \#4 and \#5.

 The surprising strength of the Na\,{\sc i} lines to
 target \#5 suggests a further favourable argument; summing over the three components, 
 one finds  W$_{\lambda}$(D2) $\approx$\,1000\,m\AA, much stronger than the  
 expected level from the ISM. Moreover, the individual D1 and D2 components, 
 which are only partially resolved our velocity resolution (6\,km\,s$^{-1}$), 
 are in fact probably saturated, and the ratio W$_{\lambda}$(D2)/W$_{\lambda}$(D1) is
 found to be 980/740\,$\sim$\,1.3, significantly less than a factor of 2, indicating that 
 the lines are not formed in the linear regime of the curve of growth. A much better
 resolution ($\sim$\,1\,km\,s$^{-1}$) would be necessary to accurately derive
 column densities, and assess, e.g. the importance of blending, saturation 
 and wing shapes. However, one can obtain a rough estimate by comparing 
 the data to IS cases described in previous studies of IS Na\,{\sc i} lines. 
 By considering the IS Na\,{\sc i} survey of Hobbs (\cite{hob78}), a 
 W$_{\lambda}$(D2) value of $\approx$\,1000\,m\AA\,\, is enormous compared 
 to usual IS lines of sight, especially when one considers the Helix 
 high galactic latitude. In Hobbs' survey, the largest D2 Na\,{\sc i} widths are at 
 500, 520 and 530 m\AA, and the corresponding $N$(Na\,{\sc i}) are 3.1, 1.3 and 7.8 
 $\times$ 10$^{+13}$\,cm$^{-2}$. In view of these numbers,
 one can reasonably estimate that to \#5 the column density
 $N$(Na\,{\sc i}) is of the order of at least $\sim$\,4\,$\times$\,10$^{+13}$\,cm$^{-2}$,
 or $\sim$\,\,20 or 40 times stronger than the expectation from the galactic layer 
 ($\sim$ 1--2\,$\times$\,10$^{+12}$\,cm$^{-2}$, Sect.~4). 
 
 To conclude, we strongly favour
 a circumstellar origin for the Na\,{\sc i} components described above.
 Finding some Na\,{\sc i} in the neutral gas of the Helix is also consistent with
 other observations on several PNs (Dinerstein \& Sneden \cite{din88}, 
 Dinerstein et al. \cite{din95}).
 In addition, we shall see in the following that further evidence for a circumstellar 
 origin is carried  by  the Ca\,{\sc ii} lines analysed below.

\subsection{Analysis of the Ca\,{\sc ii} lines}

In Fig.\,\ref{f2} are displayed the Ca\,{\sc ii} regions for  all targets. Except for 
the nucleus (\#2) that provides a generous, flat continuum, all other targets show 
broad, deep Ca\,{\sc ii} photospheric lines, implying unfortunately low $S/N$ ratios. 
Note that with the exception of \#7, the spectra have been normalised to the highest point
in the displayed range so as to best display the details, since the IS or circumstellar
features must be sought against the local continuum  noise. For \#7, the signal is much too
faint and the $S/N$ too poor for showing any detection, because the velocity range falls
in the very deep core of the Ca\,{\sc ii} lines (\#7 a K giant). For this star, 
the line profile have been shifted up by 0.2 for clarity. 

The IS Ca\,{\sc ii} line is clearly seen to target \#2, with a strength  in
very good agreement with those of the neighbouring stars of Table 2. This IS component at 
$v_{\rm helio}$\,$\sim$\,+10 km\,s$^{-1}$ is also seen, more or less clearly, to
\#1, \#8, \#4 and \#5. For  \#3, \#6 and \#7, the noise is too large.

It is therefore quite surprising to notice that, for  \#4, absolutely no trace
of a Ca\,{\sc ii} line is seen for the strong, double component at $\sim -20$\,km\,s$^{-1}$.
Similarly, it is  even more striking that, to  \#5, 
the components at $v$\,$<$\,$-5$\,km\,s$^{-1}$ , being nearly saturated in Na D, have no
Ca\,{\sc ii} counterparts. The contrast with the case of \#2 is obvious.

This lack of Ca\,{\sc ii} components can be  easily understood in the case
of absorption by circumstellar matter. There have been already a number of studies
concerning abundances in the gas phase of PN that showed that Ca is
very depleted in the ionized phase, otherwise e.g. the [Ca\,{\sc ii}] 7291\,\AA\, and 
7323\,\AA\, forbidden lines would be of great strength, whereas they are generally absent. 
The commonly accepted explanation is that, even in the ionized phase,  
Ca is almost completely 
locked in very robust grains that resist the effects of the hard UV radiation or
interaction with the PN plasma  (see for example Shields \cite{shi83}; 
Ferland \cite{fer93};  Volk et al. \cite{vol97}; Dopita et al. \cite{dop94};
Dopita et al. \cite{dop00}).
These grains are not easily directly observed (or/and their abundance measured),
 but they are necessary to explain the 
strong depletion of various refractories. This extreme Ca 
depletion in PNs is quite general and occurs in both O-rich or C-rich PNs. 
It seems to us natural to deduce that since Ca is depleted in PNs, 
e.g. {\it their ionized phase}, Ca is also very depleted in the neutral phase 
that we are probing with Na\,{\sc i} (Na is
not a refractory species and is not, or only  very lightly, depleted). 
Our interpretation is  the Ca atoms reside in 
the hard cores, condensed at high temperature, of the heterogeneous grains that 
form in the cool AGB winds and that eventually lose their more fragile 
envelope during the PN formation. 

\subsection{A search for circumstellar diffuse bands}

%%%%=====================================================================================
\begin{figure}
\resizebox{\hsize}{!}{\rotatebox{-90}{\includegraphics{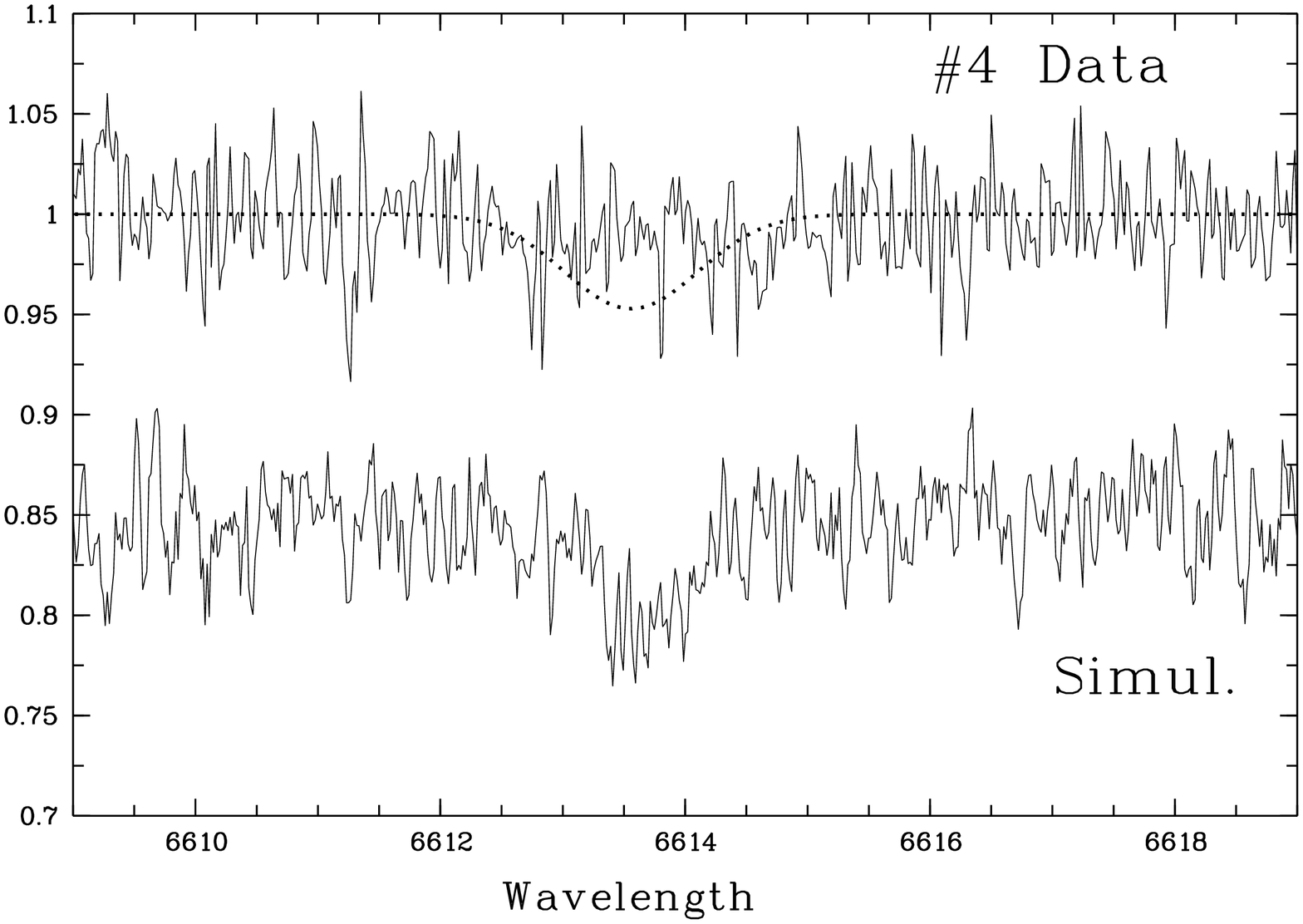}}}
\resizebox{\hsize}{!}{\rotatebox{-90}{\includegraphics{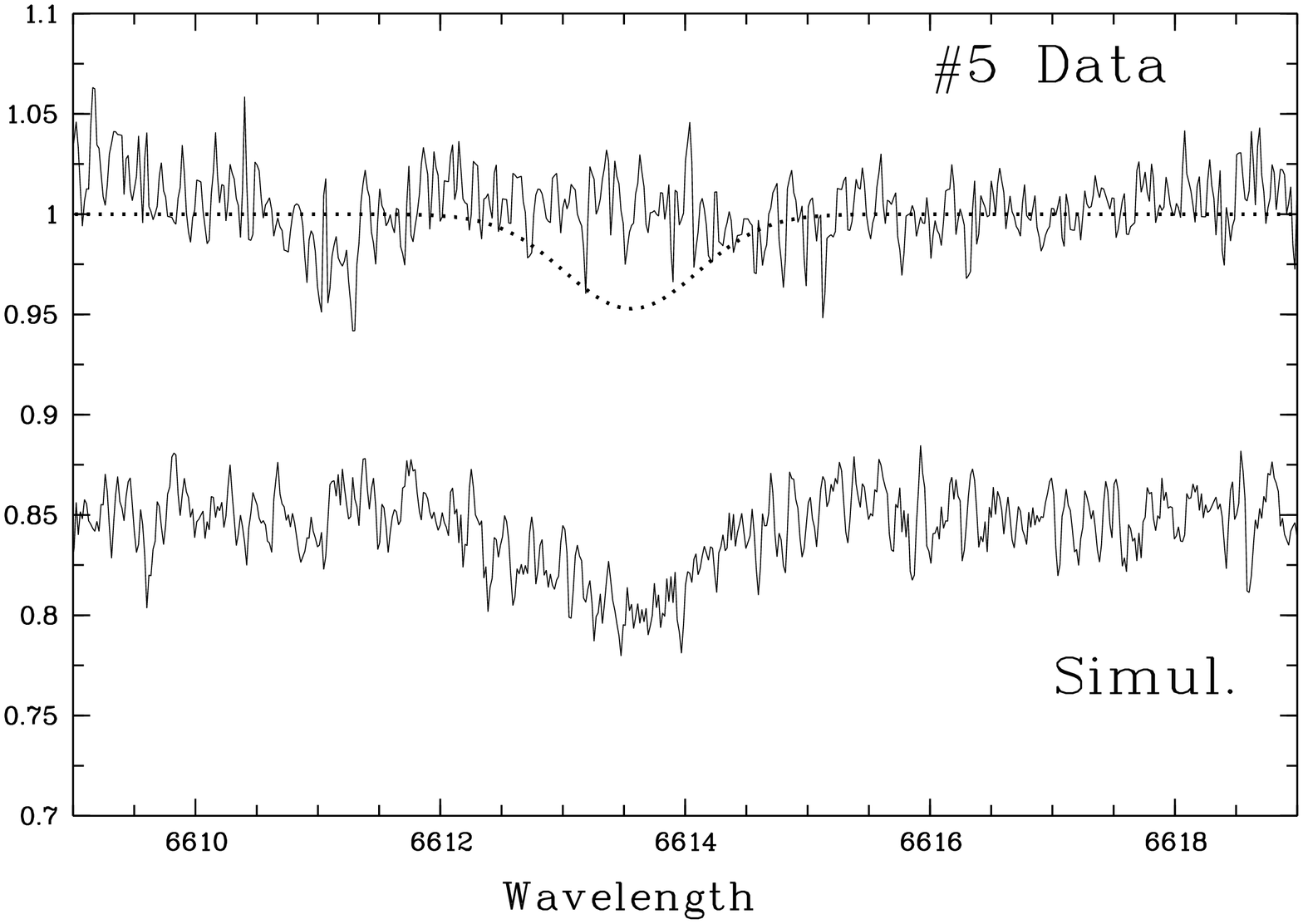}}}
\resizebox{\hsize}{!}{\rotatebox{-90}{\includegraphics{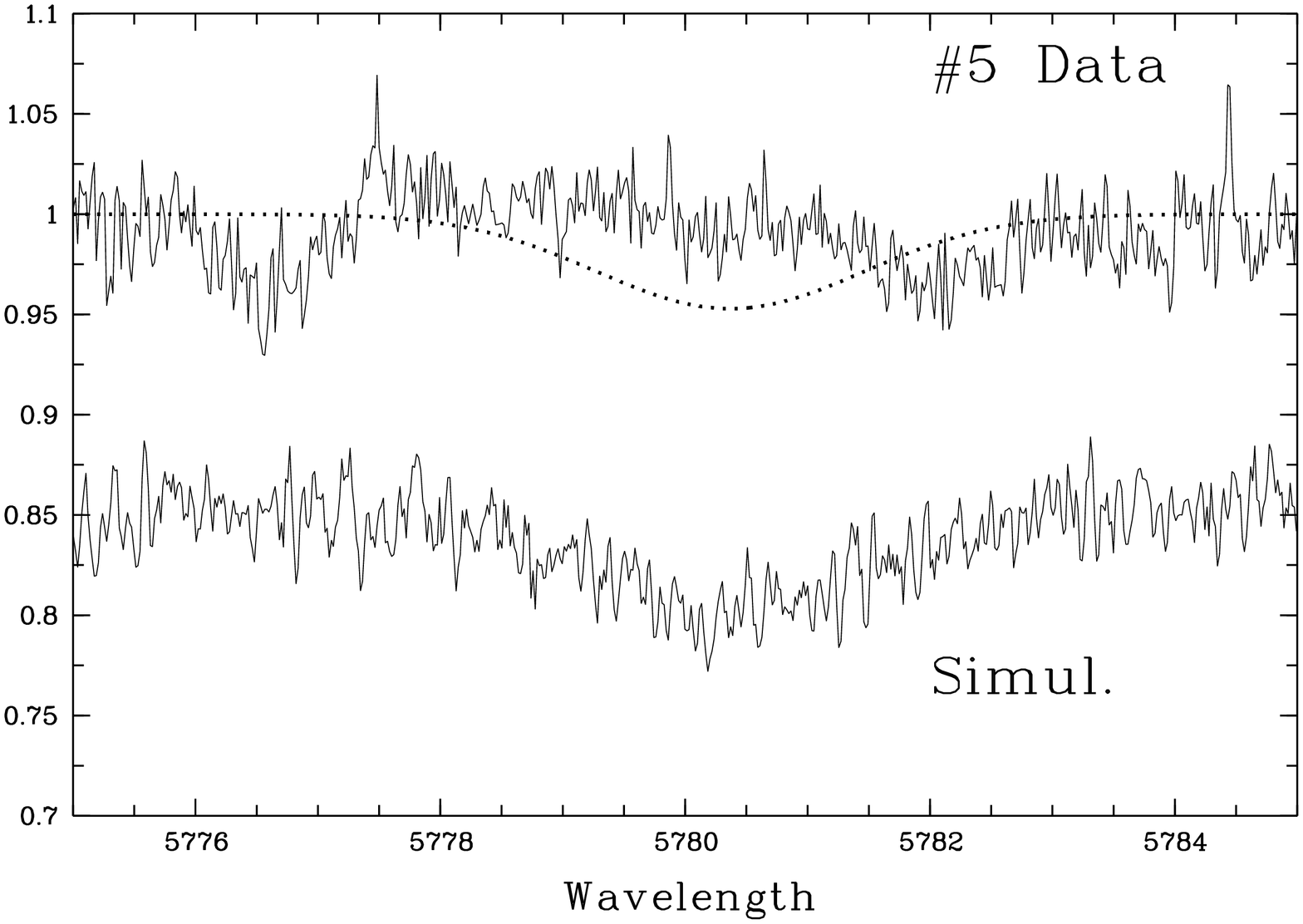}}}
\caption[]{{\it Upper panel}: Spectrum of \#4 in the region of the 6614\,\AA\, 
DIB (``Data''). The overplotted dotted line is the expected 
6614\,\AA\, feature  for an equivalent width of 50\,m\AA\, and no noise. 
The lower curve (``Simul.'') is the same DIB simulated after adding noise 
similar to the data noise. 
{\it Middle panel}: Same for \#5, for which the data $S/N$ is significantly better. 
If present, a 6614\,\AA\,  DIB of 50\,m\AA\, would be very clearly seen to \#5. 
{\it Lower panel}: Spectrum of \#5 near the 5780 DIB. 
A  DIB at 5780\,\AA\, of 100\,m\AA\, is shown and  would be fairly well detected.}
\label{f3}
\end{figure}
%%%%%%%%%%%%%%%%%%%%%%%%%%%%%%%%%%%%%%%%%%%%%%%==============================================

 We searched for DIBs in all the targets spectra, and found none. For targets
 probing the ISM (see above), this lack of DIBs is fully expected, because the
 IS excess is only $E_{\rm B-V}$\,$\sim$\,0.03, and the $S/N$ of our spectra is limited,
 despite $\sim$\,1\,hr exposure times with VLT-UVES.
 For example, from the work of Kumar (\cite{kum86}) on low colour excess stars (see his Table 1),
 one finds that, for $E_{\rm B-V}$\,=\,0.10\,$\pm$\,0.01, the 6614\,\AA~ DIB equivalent 
 width (noted $W_{\lambda}$)
 is $W_{\lambda}$(6614)\,=\,20\,m\AA\, (4 stars, $\sigma$\,=\,2\,m\AA);
 for  $E_{\rm B-V}$\,=\,0.075\,$\pm$\,0.005, $W_{\lambda}$(6614)\,=\,5\,m\AA\, 
 (8\,\,stars, $\sigma$\,=\,4\,m\AA). 
  Alternatively, Galazutdinov et al. (\cite{gal98}) show for the $\lambda$5780 DIB that
 equivalent widths of the order of 30--40\,m\AA\, are observed for their lowest colour excess
 sightlines ($E_{\rm B-V}$\,=\,0.05\,$\pm$\,0.01, comparable to those discussed above). 
 The DIB strengths due to the ISM in the 
 direction of the Helix are thus quite small (perhaps as low  as 
 $\sim$\,5\,m\AA\,  but possibly up to 30\,m\AA) and 
 simulations presented below show that our experiments cannot detect such weak  
 DIB signals.  However, our experiment may only be perhaps
 a factor 2--3 insufficiently sensitive to detect the ISM in the strongest DIBs.

 Figure \ref{f3} shows the observed spectra of targets \#4 and \#5 in the region 
 of the 6614\,\AA~ DIB (labelled ``Data''). For
 both objects, we also show  the DIB profiles that would be seen if their equivalent
 width were 50\,m\AA\, either with an infinite $S/N$ (overplotted dotted line), or
  with an added noise similar to the data (lower curve). The DIB profile
 was approximated by a Gaussian with FWHM\,1.0\,\AA\, (Walker et al. \cite{wal00}). 
 The DIB central wavelength is the rest value,
 6613.56\,\AA, as given by Galazutdinov et al. (\cite{gal00}), with uncertainty $\sim$ 0.1\,\AA. 
 No velocity shift was applied since the Earth velocity shift of +20\,km\,s$^{-1}$ 
 and the average heliocentric velocity of the Helix ($-28$\,km\,s$^{-1}$) almost cancel. 
 One can see that for \#4, a 50\,m\AA\, 6614\,\AA~  DIB would be fairly
 detectable, and for \#5 where the $S/N$ is better, it would be very clearly seen. 
 On the basis of these plots, a 3\,$\sigma$ upper limit to 
 $W_{\lambda}$(6614) for target \#5 is taken as $\sim$\,25\,m\AA. 
 
 The limits for other DIBs, e.g. at 5780\,\AA , are not better (see Fig.\,\ref{f3}). 
 The 5780\,\AA\,  DIB  is generally stronger in equivalent width 
 than $\lambda$6614, but it is also 
 twice wider; $\sim$\,2.0\,\AA\, FWHM, as observed by Galazutdinov et al. (\cite{gal98}). 
 In our spectra, its detection is not easier than for $\lambda$6614, 
 especially because of faint photospheric lines lying in this
 region. In the end, for $W_{\lambda}$(5780), we derive an
 upper limit of $\sim$\,70\,m\AA. 
  
  Instead of comparing the data with simulations, 
 one might derive  3\,$\sigma$ upper limits with the usual formula, as given by, 
  e.g.,\, Cowie \& Songaila (\cite{cow86}): for a line that falls over 
  $m$ resolution elements, the $3\sigma$ detectable equivalent width is 
 $W(3\sigma)$\,=\,3\,$\lambda$\,$m^{1/2}$\,$R^{-1}$\,$(S/N)^{-1}$, 
 where $S/N$ is the signal-to-noise ratio in each resolution element.
 For the $\lambda$6614 DIB, one finds with our data $W(3\sigma)$\,$\sim$\,10\,m\AA. 
 However, the above formula, while taking into account the observed $S/N$, assumes an absence of 
 other (photospheric) lines,
 which may not be the case (see Fig.\,\ref{f3}).  
 Therefore, we believe that the limits derived from simulations are more 
 appropriate.
 
\section{Discussion}

  The primary goal of this work was to search for circumstellar diffuse bands: none 
 were found, but we discovered two targets, \#4 and \#5, whose lines of sight indeed probe
 some neutral PN gas, with such a strong Na\,{\sc i} absorption, and strong Ca 
 depletion, that the two together cannot possibly  be due to the foreground or 
 background ISM. 

\begin{figure}
   \centering
   \resizebox{\hsize}{!}{
  {\rotatebox{-90} {\includegraphics{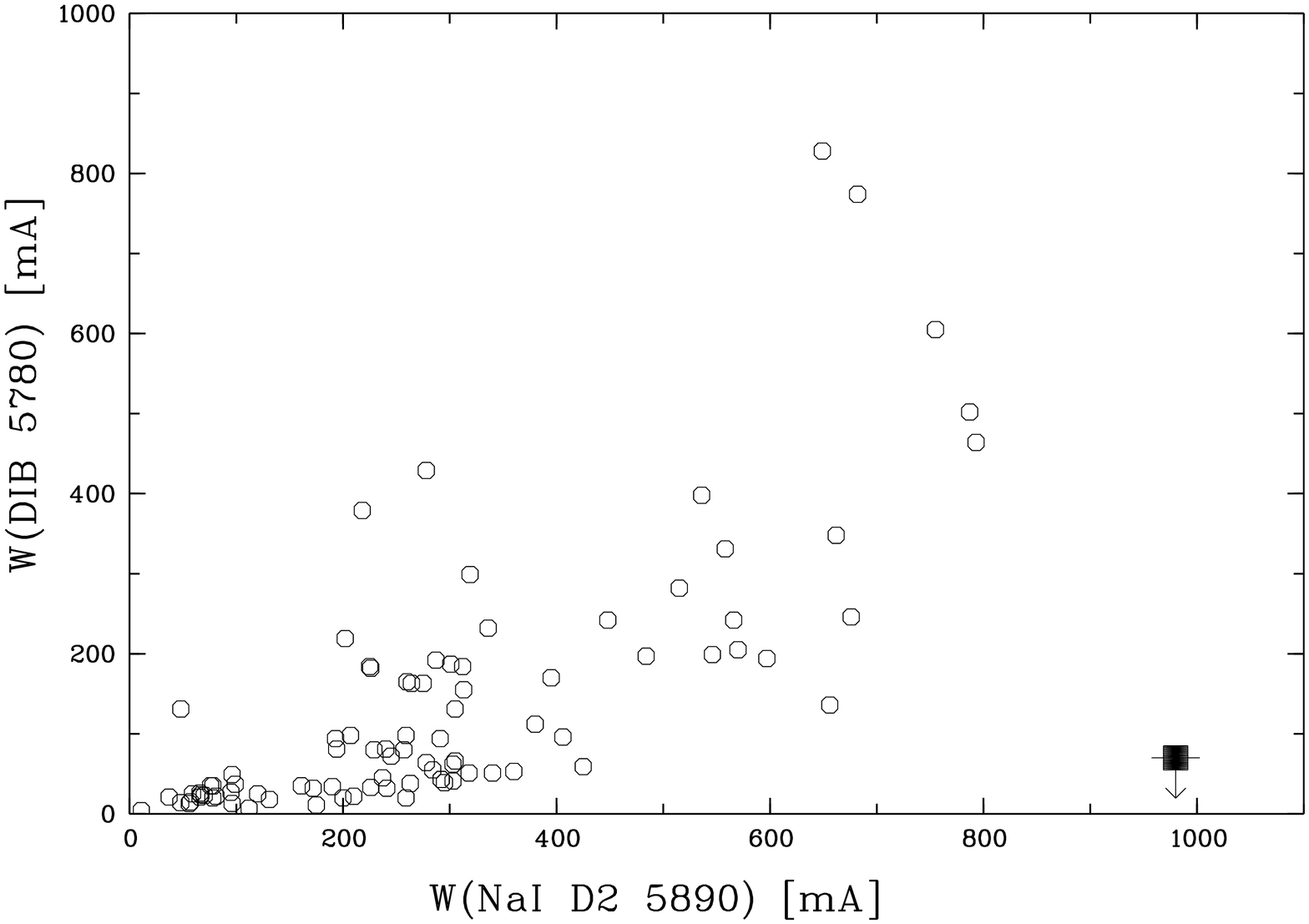}}   }
  } 
  \caption{Plot of the equivalent widths of the 5780\,\AA~DIB {\it versus} the 
   Na\,{\sc i} D2 5890\,\AA\, equivalent widths (open circles), for interstellar 
   data of Herbig (\cite{her93}). The solid square at lower right
   represents the upper limit of $\lambda$5780 for target \#5, 
   which has $W_{\lambda}$(Na\,{\sc i} D2) $\sim$ 1000\,m\AA. This plot
   illustrates the deficiency of $\lambda$5780, 
   further supporting the {\it circumstellar} 
   nature of the matter probed by this target.}
   \label{f4}
   \end{figure}

 A further indication that the sightline to \#5 is far from typically interstellar is obtained by
 plotting $W_{\lambda}$(5780) $versus$ $W_{\lambda}$(D2) (Fig.\,\ref{f4}). The data for IS
 lines were taken from the compilation of Herbig (\cite{her93}) and involves sightlines with
 $E_{\rm B-V}$ between 0 and 1.5. We see a clear increase of $W_{\lambda}$(5780)
 with increasing D2 equivalent width. The representative point
 of sightline \#5 is the full black square. Not only $W_{\lambda}$(D2) appears
 huge compared to other sightlines, especially when one realizes that \#5 is 
 at high galactic latitude, but this plot suggests that if {\em only} IS matter were
 probed, then we would see $W_{\lambda}$(5780)$\ga$\,400\,m\AA,
 which is definitely not the case. Therefore, the material seen in the line of sight to
 \#5 must be overwhelmingly of circumstellar origin.

 The fact that no DIBs have been seen is consistent with their absence in several other
 archetypal sources with carbon-rich circumstellar material. For example, in the case of
 IRC\,+10$^{\circ}$\,216 which is very rich in molecules, including carbon chains, 
 no DIB was detected on a sightline for which the Na\,{\sc i} absorption lines were
 heavily saturated, the K\,{\sc i}  lines were very strong  with 
 $W_{\lambda}$\,$\sim$\,500\,m\AA~ (Kendall et al. \cite{ken02}), 
 and the C$_2$ molecule was detected (Kendall et al.  \cite{ken04}).
 In the case of the post-AGB star HR\,4049, Waters et al. (\cite{wat89}) showed that the
 circumstellar dust is carbon-rich and displays the emission features at 7, 9 and 11.3
 $\mu$m attributed to PAHs, but no DIB was detected. Waters et al. (\cite{wat89})
  obtained a stringent 
 limit on the DIB abundance, with $W_{\lambda}(6614) < 4.2$ m\AA\, despite
 a large circumstellar reddening, $E_{\rm B-V}$ being between 0.19 and 0.29.
 Compared to these two objects, the Helix is a third interesting case; its nucleus is much
 hotter than HR\,4049, but the PN also still contains much circumstellar 
 molecular and dusty material left by the AGB progenitor wind. A peculiarity
 of the Helix is that there are no PAH mid-infrared features, interpreted by Cox et al. 
 (\cite{cox98})
 as due to a destruction of small grains by the hard radiation field of the nucleus in the
 early PN phase.
 
 Finding no DIBs in these three typical but quite different
 objects suggests that DIBs are not fabricated in AGB envelopes or PNs. However, 
 it will now become even more  necessary to $i$.) fully explain the origin of the
 emission bands in the Red Rectangle and their possible relation to DIBs; 
 and $ii$.) further examine the few cases
 of carbon stars and post-AGB objects mentioned in Sect.~1 for which evidence for 
 circumstellar diffuse bands have been put forward. 
 
 It will also be interesting  to perform specific observations of targets \#4 and \#5 
 with a higher spectral resolution and/or a better signal-to-noise ratio in order to detect 
 other species, e.g. K\,{\sc i}, C$_{2}$ or CN,  to establish whether 
 these circumstellar lines of sight probe any molecular matter, to  infer 
 the total hydrogen column density and obtain a more quantitative upper limit to the 
 DIB abundance.

%=========================== figure showing $W_{\lambda}$(5780) versus $W_{\lambda}$(NaI 5890)
%%%%%%%%%%%%%%%%%%%%%====================================================================

  \section{Summary and conclusions}
 
 1.-- With the goal of studying the neutral gas
 of the Helix PN and the presence of circumstellar
 diffuse bands, VLT/UVES  spectroscopy was achieved 
 on 8 targets lying within
 $\sim$\,13\arcmin~ of the PN centre, including its nucleus. 
 Using  available 2MASS and UBV photometry, the spectral 
 types of the targets could be estimated. This allowed the  
  derivation of  distances, which are  found to be in the
 range $\sim$ 200\,pc to 1500\,pc. Thus, several targets
 are located far beyond the Helix and can be used as background
 sources, angularly placed on or off the visible nebula.
 
 2.-- Of the 8 targets, four (including the nucleus)
  can particularly be exploited because they show little 
 blending between  the interstellar/circumstellar Na\,{\sc i} lines 
 and the photospheric lines of the background target itself, 
 which are either absent, favorably shifted by the stellar radial 
 velocity, or/and broadened by stellar rotation. 
  
 3.--  By analysing  the  extinction maps around the Helix, and  IS
 lines observed towards 3 bright stars in a $\sim$\,5\degr~zone around the Helix,
 we could estimate the expected contribution 
 of the foreground and background ISM, in terms of equivalent
 width of the Na\,{\sc i} lines (W$_{\lambda}$(D2) $\sim$\,200m\AA), 
 and the expected range of heliocentric velocities for IS lines.
 
 4.-- It appears that many of the Na\,{\sc i} components observed in the target
 spectra are probably of IS origin. However, for two targets (\# 4 and \#5), 
our analysis shows that {\it circumstellar} Na\,{\sc i} is observed in absorption. 
 For  \#5, a  very strong Na\,{\sc i} column density is derived, at least 
 20 times larger than  the expected IS contribution. 
  The circumstellar origin is also supported by the fact that no corresponding 
 absorption Ca\,{\sc ii} K line is detected, in contrast to IS lines. This fact
 is well understood: since Ca is always very depleted  in the ionized phases of
 planetary nebulae, it is probably  depleted as well and locked 
 in the circumstellar grains of the cool neutral phase, where dust has not yet been 
 subject to strong UV radiation and plasma interaction. 
 
5.-- No diffuse bands were detected. The expected bands from the foreground
and background ISM are not strong enough ($W_{\lambda}$(6614)\,$\sim$\,5 to 20\,m\AA) to 
be seen in our spectra. Nor do the circumstellar lines of sight to \#4 and \#5
show any bands. 
Qualitatively, our data do not favour abundant fabrication of diffuse band 
carriers in the Helix.

\begin{acknowledgements}
The authors warmly thank Claudio Melo, Jonathan Smoker, and the ESO staff,
for their assistance during observations. Thanks  also to Christophe Mercier
at Montpellier for his help in making Figure 1. We also thank  the anonymous referee
for comments that helped to improve the paper. N.M. acknowledges support from the
French CNRS Program  {\it Physico-chimie du Milieu Interstellaire}. T.R.K.
acknowledges support from the French {\it Minist\`{e}re de la Recherche}. The publication 
makes use of data products from 2MASS, which is a joint project of the
Univ. of Massachusetts and the Infrared Analysis and Processing Centre, California 
Institute of Technology, funded by  NASA and NSF. 
This research has also made use of the ESO MIDAS data analysis 
software, the NOAO IRAF software and the SIMBAD database, operated
at CDS, Strasbourg, France. 
    \end{acknowledgements}

\end{document}